\documentclass[12pt]{article}
\usepackage{amsmath,amssymb}
\def\R{\hbox{{\rm I}\kern-0.2em{\rm R}\kern0.2em}}
\def\R{\hbox{{\rm I}\kern-0.2em{\rm R}\kern0.2em}}
\def\s{\hbox{{\rm \subset}\kern-0.2em{\rm +}\kern0.2em}}
\def\D{\hbox{{\rm I}\kern-0.2em{\rm D}\kern0.2em}}

\def\be{\begin{equation}}
\def\ee{\end{equation}}

\def\({\left(}
\def\){\right)}
\def\[{\left[}
\def\]{\right]}
\def\bc{\begin{center}}
\def\ec{\end{center}}

\usepackage{epstopdf}
\usepackage{graphicx}
\usepackage{color}
\usepackage{float}

\begin{document}

{\large \bf Effect of quintessence on the energy of the Reissner-Nordstrom black hole}

\textit{Ibrar Hussain and Sajid Ali}

School of Electrical Engineering and Computer Science, \\
National University of Sciences and Technology,\\
H-12, Islamabad, Pakistan.\\

E-mail: ibrar.hussain@seecs.nust.edu.pk, \quad sajid$\_$ali@mail.com

{\bf Abstract}. The energy content of the Reissner-Nordstrom black hole 
surrounded by quintessence is investigated using approximate Lie symmetry 
methods. It is mainly done by assuming mass and charge of the black hole as 
small quantities ($\epsilon$), and by retaining its second power in the 
perturbed geodesic equations for such black hole while neglecting its 
higher powers. Due to the presence of trivial second-order approximate 
Lie symmetries of these perturbed geodesic equations, a rescaling of 
the geodetic parameter gives a rescaling of the energy in this black hole. 
Interestingly we obtain an explicit relation of the rescaling factor 
that depends on the square of the charge to mass ratio of the black hole, 
the normalization factor $\alpha$, which is related to the state parameter 
of the quintessence matter, and the coordinate $r$. A comparison 
of this rescaling factor with that of the Reissner-Nordstrom black hole 
(Hussain et. al SIGMA, 2007), without quintessence is given. It is observed 
that the presence of the quintessence field reduces the energy in this 
black hole spacetime. Further it is found that there exists a point outside the 
event horizon of this black hole where the effect of quintessence balances
the energy content in this black hole without quintessence, and where the total 
energy of the underlying spacetime becomes zero.


\section{Introduction}
To give a well accepted definition of energy is a long standing
problem in general relativity (GR) since the time of Einstein
(for detail one may see \cite{MTW}). To resolve this problem
many scientists have given their own notions of energy in GR \cite{Sab1}. These
attempts to define energy in GR include several pseudo-tensors
and approximate symmetry approaches. A review of these approaches is 
available in the literature \cite{MS1, IH1, IH2}. The pseudo-tensors are coordinate 
dependent quantities and therefore violates the basic principle of GR.
On the other hand approximate symmetry approaches, except the approximate 
Lie symmetry approach have their own drawbacks which are pointed out in \cite{IH2}. The
approximate Lie symmetry method \cite{NH1}, to define energy in GR is
comparatively new and free of the drawbacks.

The approximate Lie symmetry methods for differential equations \cite{NH1},
have been applied to the approximate (perturbed) geodesic equations of
different gravitational wave spacetimes and some black hole spacetimes
of GR and other theories of gravity \cite{IH2, IH3, MS2, MS3, IH4}. For all these spacetimes energy
rescaling factors have obtained. These spacetimes include plane-fronted and
cylindrical gravitational waves \cite{IH5, IH6}, Reissner-Nordstrom \cite{IH1}, Kerr-Newman \cite{IH7},
Kerr-Newman AdS \cite{IH8}, BTZ \cite{IH3}, Bardeen \cite{MS2}, stringy black hole \cite{MS3}
and a slowly rotating black hole in the Horava-Lifshitz theory of gravity \cite{IH4}.

The evidence for the accelerated expansion of our Universe is supported by some
Cosmological observations like the Supernovae Ia, the Cosmic Microwave Background
radiation anisotropies and X-ray experiments \cite{Pe, PB, DN}. Astrophysicists and cosmologists
consider a missing energy component with negative pressure called dark energy,
responsible for this accelerated expansion of the Universe. There are different
candidates for dark energy (see e.g., \cite{Mir}). One of these is known as quintessence.
This is defined as a scalar field with the equations of negative state parameter,
which is the ratio of the pressure and density \cite{LZ}. In 2003,
Kiselev derived a charged black hole solution with quintessence term, of
the Einstein field equations (EFEs) \cite{Kis}. This solution reduces to the
Reissner-Nordstrom black hole solution of the EFEs in the absence of the
quintessence term. The energy expression (rescaling factor) in the case of
Reissner-Nordstrom black hole was obtained in \cite{IH1} via approximate Lie
symmetry methods. In this paper we are interested to apply the approximate Lie
symmetry methods to the Reissner-Nordstrom black hole with quintessence to
look at its enrgy content and compare it with that of the Reissner-Nordstrom black hole.
In particular we want to study the effect of the quintessence term in the energy of this
black hole.

This paper is organized as follows. In the next Section mathematical
definitions to be utilize are given. In Section 3 we will discuss the exact,
first-order and the second-order approximate symmetries of the geodesic equations
of the perturbed Reissner-Nordstrom black hole surrounded by quintessence. In
the same Section we will derive the energy rescaling factor for this black hole.
In Section 4 the effect due to the presence of quintessence field is discussed in detail. 
A summery of the work done here is given in the last Section.

\section{Basic Definitions}

Under a point symmetry transformation \cite{NH1, NH2},
\begin{equation}
{\bf V}=\zeta  (\tau ,x^\alpha )\frac{\partial}{\partial \tau }+\psi ^\beta (\tau ,x^\alpha)\frac{\partial}{\partial x^\beta},
\end{equation}
where $\alpha , \beta = 0,1,2,3$, a second-order approximate Lie symmetry is a vector field
\begin{equation}
{\bf V}={\bf V}_0+\epsilon {\bf V}_1+\epsilon^2{\bf V}_2+O(\epsilon^3),
\end{equation}
for a system of perturbed ordinary differential equations (here the system of geodesic equations)
\begin{equation}
{\bf G}={\bf G}_0+\epsilon {\bf G}_1+\epsilon^2{\bf G}_2+O(\epsilon^3),
\end{equation}
if the following condition holds
\begin{equation}
({\bf V})({\bf G})_{{\bf G}=O(\epsilon^3)}=O(\epsilon^3).
\end{equation}
In (2) ${\bf V_0}$ is the exact (when $\epsilon=0$) symmetry generator for the system of
the exact geodesic equations ${\bf G_0}$. The exact Lie symmetry can be determined from
\begin{equation}
({\bf V}_0)({\bf G}_0)_{{\bf G}=0}=0.
\end{equation}
The vector fields ${\bf V}_1$ and ${\bf V}_2$ are the first-order and
second-order approximate parts of the approximate symmetry generator ${\bf V}$,
respectively. Similarly ${\bf G}_1$ and ${\bf G}_2$ are the first-order and second-order
perturbed parts of the system of geodesic equations ${\bf G}$, respectively. Since
the geodesic equations are second order ordinary differential equations, therefore
the second prolongation of the vector fields ${\bf V}_0$, ${\bf V}_1$ and ${\bf V}_2$ should
be used in (4) and (5), which is available (for example) in \cite{IH2}. The second-order
approximate symmetry generator is said to be nontrivial approximate symmetry generator if it
is not proportional to any one of the lower-order (i.e exact or first-order) symmetry
generators. The second-order approximate symmetry generator is also called nontrivial if
at least one of the lower-order symmetry generators are nonzero for it. In the case of
trivial approximate symmetry generators it is also possible that the lower-order
symmetry generator cancel out in the set of determining equations which can be obtained
from (4). It is worth remarking that the interesting result of energy rescaling (explained in the next
Section) comes from the applications of the perturbed system of geodesic equations
in the subscript of (4), as required (for more detail see \cite{IH2}).

The first prolongation of the vector field (1) is
\begin{equation}\label{9}
\mathbf{V}^{[1]}=\mathbf{V}+(\dot \psi ^\alpha_{,\tau } + \gamma^\alpha_{,\beta} \dot{x}^\beta - \zeta _{,\tau }\dot{x}^\alpha
-\zeta _{,\beta }\dot{x}^\beta  \dot{x}^\alpha)  \frac{\partial }{\partial \dot{x}^\alpha }.
\end{equation}
The vector field ${\bf V}$ is called a Noether gauge symmetry generator for the Lagrangian
$L(\tau ,x^\beta , \dot{x}^\beta )$,
if the following condition satisfies
\begin{equation}
\mathbf{V^{[1]}}L+(\mathbf{d} \zeta )L=\mathbf{d}  g.
\end{equation}
Here $g(\tau ,x^\beta )$ is a gauge function and the total derivative operator $\bf{d}$
is given by
\begin{equation}\label{12}
\mathbf{d} =\frac{\partial }{\partial \tau }+\dot{x}^{\beta} \frac{\partial
}{\partial x^{\beta} }.
\end{equation}
Throughout this article the dot stands for derivative with respect to the geodetic
parameter $\tau $ and Einstein summation convention is assumed. The following 
Noether theorem \cite{NT1}, reveals the significance
of Noether symmetries.

{\bf Theorem 1}. If {\bf V} is a Noether gauge symmetry generator
corresponding to a Lagrangian $L(\tau ,x^{\beta },\dot{x}^{\beta })$ of the
Euler-Lagrange equations of motion, then
\begin{equation}
I={\zeta }L+(\psi ^{\beta }-\dot{x}^{\beta }{\zeta })\frac{\partial L}{\partial
\dot{x}^{\beta }}- g, \label{6}
\end{equation}
is a constant of motion associated with the symmetry generator {\bf V}.
The proof of this theorem can be seen for example in \cite{LV}.

\section{Approximate Lie Symmetries and the Energy content of the Reissner-Nordstrom
black hole surrounded by quintessence}

The line element for the static charged black hole surrounded by quintessence is \cite{Fer}
\begin{equation}
ds^2 = f(r) dt^2 - \frac{dr^2}{f(r)} - r^2 d \Omega^2 . \label{line-1}
\end{equation}
where $d\Omega$ is the solid angle defined by $ d\Omega^2 = d\theta^2 +\sin^2{\theta} d\phi^2$ and
 $f(r)$ is a function  of the form
\begin{align}
f(r) = 1 - \frac{2M}{r}+ \frac{Q^2}{r^2}-\frac{\alpha} {r^{3{\omega_q}+1}}. \label{form-1}
\end{align}
In (11) $M$ is the mass and $Q$ is the total charge of the black hole. The factor $\alpha$
is related with the state parameter $\omega_q$ by
\begin{equation}
\alpha=-\frac{2\rho{_q}r^{3(\omega_q+1 )}}{3\omega_q},
\end{equation}
where $\rho_q$ is the energy density of the quintessence matter. The pressure $p_q$
for the quintessence matter has the relation with $\rho_q$ as
\begin{equation}
p_q=\omega_q \rho_q.
\end{equation}
The range of $\omega_q$ for quintessence matter is given in the literature (see e.g.,\cite{Fer}).
In this paper we take $\omega_q = -\frac{2}{3}$, for which the charged black hole with
quintessence, given by (10) and (11), becomes the simplest nontrivial case \cite{Fer}.
For this value of $\omega_q$ the function in (11), takes the following form
\begin{equation}
f(r) = 1 - \frac{2M}{r}+ \frac{Q^2}{r^2}-\alpha r,
\end{equation}
in which case the black hole is non-asymptotically flat.

To discuss the approximate symmetries of the charged-quintessence Reissner-Nordstrom 
black hole, we assume both of its mass and charge to correspond to first and 
second order perturbation of $\epsilon$, i.e.,
\begin{equation}
2M=\epsilon,\quad Q^2=k\epsilon^2.
\end{equation}
To avoid the naked singularity \cite{MTW}, we must have
\begin{equation}
M^2\geq Q^2,
\end{equation}
therefor $0<k\leq \frac{1}{4}$. Equation (14) now becomes
\begin{equation}
f(r)=1-\alpha r-\frac{\epsilon}{r}+\frac{\epsilon^2}{r^2}.
\end{equation}

By retaining the second power of $\epsilon$ and neglecting its higher powers in the perturbed geodesic equation of
this spacetime we get
\begin{align}
&
\ddot{t} = - \frac{\alpha \dot{r}\dot{t}}{\alpha r -1} + \epsilon \frac{(2\alpha r-1) \dot{r}\dot{t}}{r^2(\alpha r -1)^2} - \epsilon ^2 \frac{ \left ( 6k\alpha^2 r^2 - 2\alpha \left (5k-2\right )r + 2(2k-1)\right ) \dot{r}\dot{t}}{r^3(\alpha r -1)^3} + \mbox{O}(\epsilon^3)
, \nonumber \\
&
\ddot{r} =  \frac{ \alpha ( \dot{r}^2 - (\alpha r -1 )^2 \dot{t}^2)}{2(\alpha r -1)} +r(1-\alpha r)(\dot{\theta}^2+\sin^2\theta \dot{\phi}^2)+
\epsilon \Big(\frac{(1-2\alpha r) \dot{r}^2 - (\alpha r -1)^2\dot{t}^2}{2r^2 (\alpha r -1)^2}\nonumber\\
&
-(\dot{\theta}^2+\sin^2\theta \dot{\phi}^2) \Big)+
 \frac{\epsilon^2 }{r^3 (\alpha r -1)^3}\Big(-k\alpha^4 r^4 \dot{t}^2 + (5k+1) \alpha^3 r^3 \dot{t}^2 +
3\alpha^2 r^2 (k\dot{r}^2 - (3k+1)\dot{t}^2) +
\nonumber \\
&
\alpha r ((7k+3)\dot{t}^2 - (5k-2)\dot{r}^2)
-(2k+1)\dot{t}^2 + (2k-1) \dot{r}^2 +\frac{k}{r}(\dot{\theta}^2+\sin^2\theta \dot{\phi}^2) \Big)
+
\mbox{O}(\epsilon^3) ,\nonumber \\
&
\ddot{\theta} = \sin{\theta}\cos{\theta} \dot{\phi}^2 - \frac{2}{r} \dot{r}\dot{\theta}, \nonumber \\
&
\ddot{\phi} = -\frac{2}{r} \dot{r}\dot{\phi} - 2\cot{(\theta)} \dot{\theta}\dot{\phi}. \label{geo-1},
\end{align}
where the last two equations are the same as for the unperturbed case because 
there was no perturbation in the solid angle. In order to investigate the 
second-order approximate symmetries of the above system we first need to 
obtain the exact and the first-order approximate symmetries of the exact and the first-order 
perturbed system of geodesic equations. For the exact case we substitute
$\epsilon =0$, in the above system (18) and get six Lie symmetry generators  from (5)
\begin{align}
&
\textbf{V}_{0} =  \tau\frac{\partial}{\partial \tau },~\textbf{V}_{1} =  \frac{\partial}{\partial \tau },~ \textbf{V}_{2} = \frac{\partial}{\partial t }, ~\textbf{V}_{3} = \frac{\partial}{\partial \phi },\nonumber \\
&
\textbf{V}_{4} =  \sin{\phi}\frac{\partial}{\partial \theta } + \cot {\theta}\cos {\phi} \frac{\partial}{\partial \phi }, \quad \textbf{V}_{5} =  -\cos{\phi}\frac{\partial}{\partial \theta } + \cot {\theta}\sin {\phi} \frac{\partial}{\partial \phi }.
\end{align}
In the above symmetry generators ${\bf V}_2$, ${\bf V}_3$, ${\bf V}_4$, and ${\bf V}_5$,
are the Killing vectors for the underlying spacetime, which correspond to energy, 
azimuthal angular momentum and angular momentum conservation. The four Killing
vectors together with the symmetry generator ${\bf V}_1$ are the
Noether symmetries, i.e., symmetries of the corresponding Lagrangian for the geodesic
equations. The corresponding conserved quantities or first integrals of the equations 
of motion can be obtained by using the Noether theorem given in Section 1.  The 
symmetry generator given by ${\bf V}_0$, is the proper Lie symmetry generator for 
the spacetime under consideration that correspond to rescaling transformation 
of the proper time $\tau$.

Using these six exact Lie symmetry generators given in (19), in the definition 
of first-order approximate Lie symmetry conditions (which can be obtained by 
retaining only the first power of $\epsilon$ in (3) and (4)) we obtain the 
same six symmetry generators which are trivial first-order approximate 
symmetry generators. Now the second-order approximate symmetry generators can 
be obtained by retaining the second-order perturbed term $\epsilon$ and neglecting
higher-order terms in the perturbed geodesic equations (18), and by employing the six 
symmetry generators given in (19) as the exact and the first-order approximate symmetries 
generators. From the solution of (4), it turns out that we do not get any new 
symmetry generator for the second-order perturbation and retain those six trivial
symmetry generators. Therefore the six exact symmetry generators given in (19) 
are obtained as trivial second-order approximate symmetries generators.

The trivial second-order approximate symmetry generators ${\bf V}_0$
and ${\bf V}_1$ in (19), correspond to
\begin{equation}
\zeta =a_0+a_1\tau,
\end{equation}
It is observed that in the set of determining equations for the first-order 
approximate symmetries of the first-order perturbed geodesic equations the 
terms involving $\zeta_\tau=a_1$ cancel out. Whereas in the case of second-order 
approximate symmetries of the second-order perturbed geodesic equations 
for Reissner-Nordstrom spacetime with quintessence the terms involving
$\zeta_\tau=a_1$, do not cancel automatically. Interestingly these collect
a scaling factor to cancel out for consistency of the determining
equations . In which case we get the scaling factor
\begin{equation}
\Big[(1-2k)+\alpha r(5k-2) -3k\alpha^2 r^2 \Big ].
\end{equation}
The time translation is related with energy conservation and $\zeta$ is 
the coefficient of $\partial /\partial \tau$ in the point symmetry 
transformation (1), where $\tau$ is the proper time. The scaling factor 
(21) corresponds to the rescaling of energy in the Reissner-Nordstrom
spacetime surround by quintessence. By substituting $k=\frac{Q^2}{4M^2}$, 
in (21) we get 
\begin{equation}
\Big[\Big (1-\frac{Q^2}{2M^2}\Big ) +  \alpha r \Big(\frac{5Q^2}{4M^2}-2 \Big) -\frac{3Q^2}{4M^2}\alpha^2 r^2 \Big ].
\end{equation}
By putting $\alpha=0$, (22) reduces to the same scaling factor which
was obtained for the Reissner-Nordstrom spacetime without quintessence 
\cite{IH1}. We see that like the Reissner-Nordstrom spacetime \cite{IH1}, 
this scaling factor involves the second power of ratio between $Q$ and 
$M$. Therefore it relates the electromagnetic self-energy to the gravitational
self-energy.

\section{Effect of quintessence}

The pressure $p_q$ of the quintessence field which is assumed to be responsible for
the accelerated expansion of the Universe is negative \cite{Pe}. It is
evident from equation (13), that $\rho_q$ has to be positive because $\omega_q$ is negative.
Therefore, from equation (12) we see that $\alpha$ is positive. Thus the energy in the 
field of Reissner-Nordstrom black hole surrounded by quintessence is differ by 
\begin{equation}
E_{Q}(r) = \alpha r(5k-2) -3k\alpha^2 r^2,
\end{equation}
from the energy in the field of Reissner-Nordstrom black hole without 
quintessence \cite{IH1}. It may be pointed that due to the presence of 
quintessence the energy of black hole varies in the radial direction 
and has at most quadratic dependence on $r$. We now investigate the 
variation in the energy of such a black hole. Notice that the presence
of the quintessence field will enhance the energy content of this black
hole spacetime if $E_{Q}$ in (23), is positive. This possibility can be
ruled out easily by studying the graph of function $E_{Q}$. The 
function increases for all values of $r$ in the range 
\begin{equation}
r < \frac{5k-2}{3k\alpha}~.
\end{equation}
Since $0<k\leq 1/4$ and $\alpha>0$, therefore the quantity on the right hand side
of (24) is always negative. This leads to a contradiction because $r$ denotes radial
distance and cannot be negative. Thus the enhancement of the energy content is not
possible of the Reissner-Nordstrom black hole surrounded by quintessence, 
due to the presence of the quintessence field. The function $E_{Q}$, attains a maximum
value at 
\begin{equation}
r = \frac{5k-2}{6k\alpha} < 0 \quad \Big(\mbox{for all} \quad 0<k\leq \frac{1}{4}\Big)
\end{equation}
and then decreases with a negative slope 
\begin{equation}
E_{Q}^{~\prime} = \alpha (5k-2) -6k\alpha^2 r <0,
\end{equation}
for all values of $r$ in the range 
\begin{equation}
\frac{5k-2}{6	k\alpha}< r <\infty ~.
\end{equation}
Therefore we conclude that in the above region the contribution due to quintessence
term $E_{Q}$, is to decrease the energy in the spacetime of this black hole. 
This supports the idea of mass reduction of black holes by the accretion of 
dark energy on black holes \cite{Bch, MJ1, MJ2}. 

Finally we find the value of $r$ at which the term due to quintessence
balances the energy content in the absence of this term.
This can be done by equating the quintessence term $E_{Q}$ to the negative
of energy term $(1-2k)$, that corresponds to the energy without 
quintessence. We get two values of $r$, namely
\begin{equation}
r_{(\pm)} = \frac{(5k-2) \pm \sqrt{4-8k+k^2}}{6k\alpha}~.
\end{equation}
The discriminant is positive for $0<k\leq 1/4$ which gives the first 
root $r_{(+)}$ as a positive quantity, i.e., $r_{(+)}>0$. On the other 
hand $r_{(-)}<0$, which is not a feasible solution. Therefore, at the 
radial distance $r=r_{(+)}$, the effect of quintessence balances the energy 
content of the black hole without quintessence. While for $r > r_{(+)}$, the
effect due to the quintessence term goes on decreasing. Figure 1 illustrates
the behavior of $E_{Q}$, for different values of $r$.
\begin{figure}[H]
\centerline{\includegraphics[width=12cm]{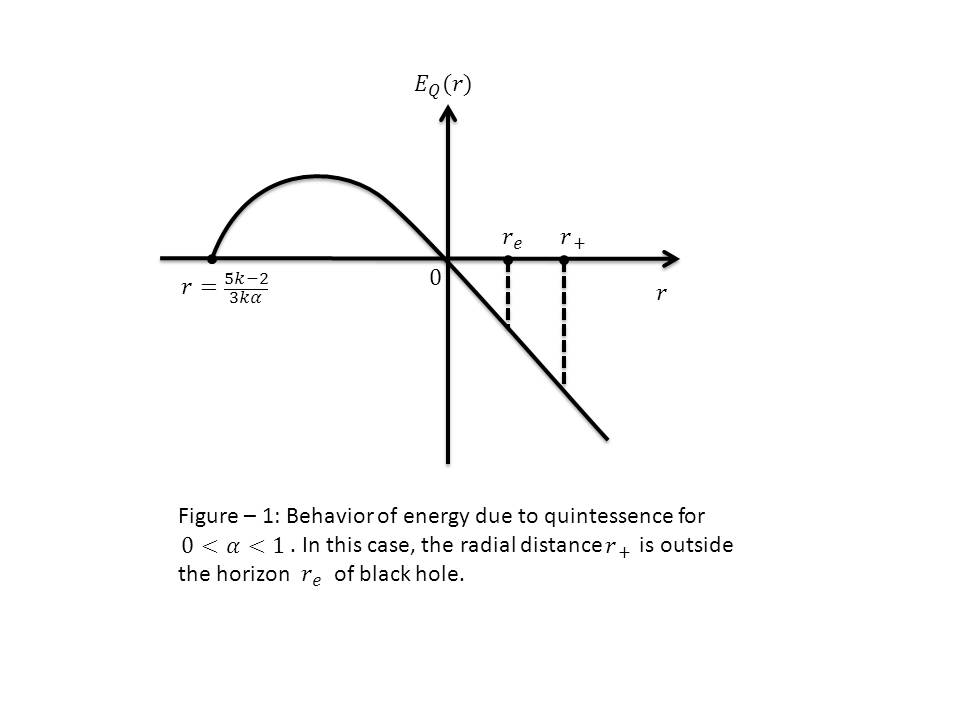}}
\end{figure}
\begin{figure}[H]
\centerline{\includegraphics[width=12cm]{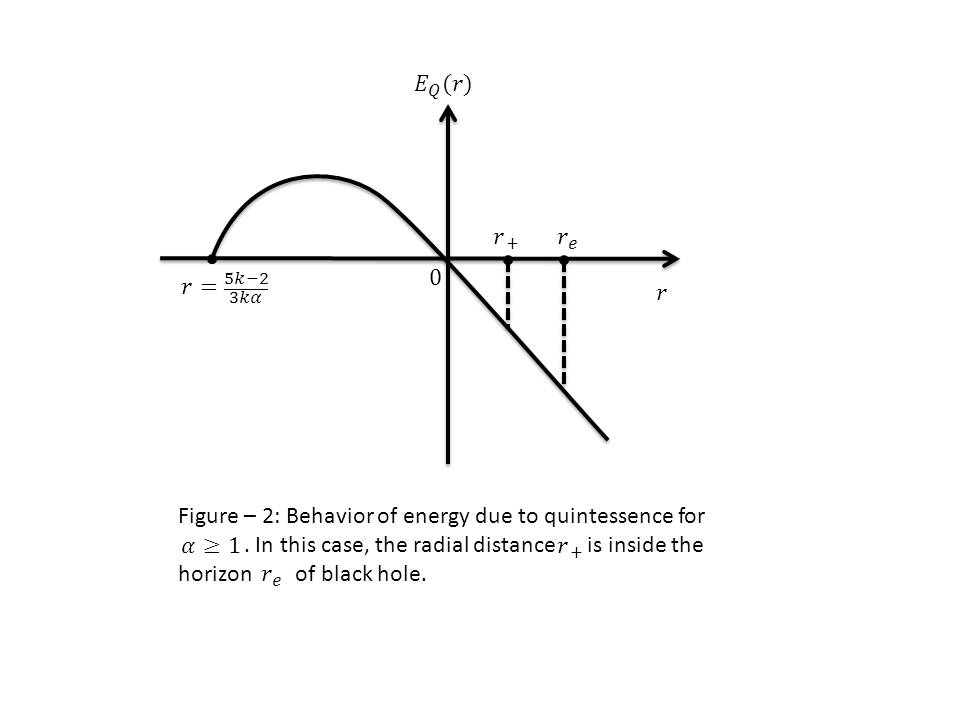}}
\end{figure}

It would be interesting to locate the point $r_{(+)}$, and check if
it lies inside or outside the event horizon of the black hole. A detailed
discussion on the event horizons of the Reissner-Nordstorm black hole 
surrounded by quintessence is given in \cite{Fer}. We consider the case in
which Reissner-Nordstorm black hole with quintessence has a single event 
horizon ($r_e$), that correspond to a single real root of $f(r)$, in the metric. 
In this case equation for event horizon $f(r)=0$, is a cubic equation in $r$ 
and has three roots in general, one in the real and two on the complex plane. 
Since the complex roots leads to naked singularities therefore violate Roger
Penrose's cosmic censorship hypothesis. Thus, the only possible real root 
correspond to a real singularity with a single event horizon. By taking 
suitable values of $Q$ and $M$, satisfying $M^2\geq Q^2$, it can easily 
be verified that (1) if $0<\alpha<1$, then the point $r=r_{+}$ lies outside 
the event horizon and (2) for $\alpha \geq 1$, the point $r=r_{+}$ lies 
inside the event horizon. Therefore we obtain critical bounds on the parameter 
$\alpha$, in the Reissner-Nordstorm black hole surrounded by a quintessence 
field. The case $0<\alpha<1$, is important because there exists a point 
outside the event horizon of black hole with quintessence where the 
energy of this spacetime vanishes completely. The other case $\alpha \geq 1$, 
brings $r_{+}$ inside the event horizon of the black hole (see Figure 2) where fundamental 
laws of Physic fail to exist and nothing can be said about it.

\section{Summary}

In this paper we have investigated the energy content of the Reissner-Nordstrom black
hole surrounded by quintessence field, by using second-order
approximate Lie symmetries. For this purpose we have considered mass $M$ and charge
$Q$ of the black hole as small parameter $\epsilon$. To determine the second-order
approximate symmetries of the second-order perturbed geodesic equations we have first studied
the exact (when $\epsilon=0$) and the first-order approximate (when second and higher 
powers of $\epsilon$ are neglected) symmetries. In the exact case we have found 
the six dimensional Lie algebra with generators given in (19). In the first-order 
and second-order approximate cases we have no non-trivial approximate Lie symmetry 
generator. We have recovered the six exact symmetry generators as trivial first-order
and second-order approximate symmetry generators.

From the application of the definition of the second-order approximate Lie symmetries
we have obtained an energy rescaling factor (22) for the Reissner-Nordstrom black
hole surrounded by quintessence. This energy rescaling factor depends on the ratio
of the charge $Q$ and mass $M$ of the black hole and relates to the electromagnetic
self-energy to the gravitational self-energy. These two parameters $M$ and $Q$
of the black hole appears quadratically in the energy rescaling factor (22).
Further this rescaling factor for the Reissner-Nordstrom black hole with quintessence
term also involve $r$ and $\alpha$. The comparison of this rescaling factor with
that for the Reissner-Nordstrom black hole \cite{IH1}, was given in Section 3. By
taking $\alpha=0$ in (22), one can recover the energy rescaling factor for the 
Reissner-Nordstrom black hole \cite{IH1}. The energy rescaling factor for the 
Reissner-Nordstrom black hole \cite{IH1}, do not depends on the coordinate $r$, 
while the factor (22) obtained here for the black hole in the presence of the quintessence 
matter involves $r$, which appears as a multiple of $\alpha$ and does 
not appear separately. Taking $\alpha=0$, the $r$ dependent terms will also 
disappear from the rescaling factor (22). Here it is observed
that the presence of the quintessence field may {\it reduce} the energy in the 
spacetime field of the Reissner-Nordstrom black hole surrounded by quintessence.
Besides, we have obtained critical bounds on the value of the normalization 
factor $\alpha$ i.e it lies between 0 and 1. It is also seen that outside the 
event horizon of the Reissner-Nordstrom black hole surrounded by quintessence,
there exist a point at which the effect due to the quintessence term balances
by the energy term of the black hole without the quintessence. The effect of 
quintessence then decreases beyond that point.

\section*{Acknowledgments} IH would like to thank TWAS-UNESCO for awarding Associateship 
at Kavli Institute for Theoretical Physics, Chinese Academy of Sciences, Beijing, China 
where some part of the work was completed.


\begin{thebibliography}{}

\bibitem {MTW} C. W. Misner, K. S. Thorne, and J. A. Wheeler, Gravitation, (W.H. Freeman, San Francisco, 1973).

\bibitem{Sab1} L. B. Szabados, Living Rev. Relativ. {\bf 7}, 4 (2004).

\bibitem{MS1} M. Sharif, Nuovo Cimento B {\bf 118}, 669 (2003).

\bibitem{IH1} I. Hussain, F. M. Mahomed and A. Qadir, SIGMA {\bf 3}, 115 (2007).

\bibitem{IH2} I. Hussain, Definition of Energy in General Relativity via Approximate Symmetries, 
(LAP Lambert Academic Publishing, 2012).

\bibitem {NH1} N. H. Ibragimov, Elementary Lie group Analysis and Ordinary Differential Equations,
(Wiely, Chichester, 1999).

\bibitem{IH3} I. Hussain, Phys. Scripta, {\bf 83}, 055002 (2011).

\bibitem{MS2} M. Sharif, and S. Waheed, Can. J. Phys. {\bf 88}, 833 (2010).

\bibitem{MS3} M. Sharif, and S. Waheed, Phys. Scr. {\bf 83}, 015014 (2011).

\bibitem{IH4} I. Hussain, J. Koer. Phys. Soc. (to appear).

\bibitem{IH5} I. Hussain, F. M. Mahomed, and A. Qadir,  Phys. Rev. D
{\bf 79} 125014 (2009).

\bibitem{IH6} I. Hussain, and A. Qadir, Nuovo Cimento B {\bf 122}, 593 (2007).

\bibitem{IH7}  I. Hussain, F. M. Mahomed, and A. Qadir, Gen. Relativ.
Grav. {\bf 41}, 2399 (2009).

\bibitem{IH8} I. Hussain, Gen. Relativ. Grav. {\bf 43} 1037 (2011).

\bibitem{Pe} S. Perlmutter et al., Astrophys. J. {\bf 517}, 565 (1999).

\bibitem{PB} P. de Bernardis et al., Nature, {\bf 404}, 955 (2000).

\bibitem{DN} D. N. Spergel et al., Astrophys. J. Suppl. {\bf 148}, 175 (2003).

\bibitem{Mir} W. Miranda, S. Carneiro, C. Pigozzo, JCAP 1407, 043 (2014).

\bibitem{LZ} I. Zlatev, L. -M. Wang and P. J. Steinhardt, Phys. Rev. Lett. {\bf 82}, 896 (1999).

\bibitem{Kis} V V Kiselev, Class. Quantum Grav {\bf 20}, 1187 (2003).

\bibitem {NH2} N. H. Ibragimov, A. H. Kara  and F. M. Mahomed,
Lie-B\"acklund and Noether Symmetries with Applications,
Nonlinear Dynamics {\bf 15}, 115 (1998).

\bibitem {NT1} E. Noether, Math. Phys.Kl. 2 (1918) 235. (English translation in transport theory and
Statistical Physics 1 (1971)) 186.

\bibitem {LV} L. V. Ovsiannikov, Group Analysis of Differential Equations, (New York: Academic
Press, 1980).

\bibitem{Fer} S. Fernendo, Gen. Rel. Grav. {\bf 45}, 2053 (2013).

\bibitem{Bch} E. Babichev, V. Dokuchaev, and Yu. Eroshenko, Phys. Rev. Lett. {\bf 93}, 021102 (2004).

\bibitem{MJ1} M. Jamil, Eur.  Phys. J. C {\bf 62}, 609 (2009).

\bibitem{MJ2} M. Jamil and A. Qadir, Gen. Rel. Grav. {\bf 43}, 1069 (2011).

\end{thebibliography}
\end{document}